\documentclass[letterpaper,twocolumn,10pt]{article}
\usepackage{times}
\usepackage{fullpage}
\usepackage{cite}
\usepackage{amsmath,amssymb,amsfonts}
\usepackage{algorithmic}
\usepackage{graphicx}
\usepackage{textcomp}
\usepackage{xcolor}
\usepackage{subcaption}
\usepackage{hyperref}
\usepackage{relsize}
\def\BibTeX{{\rm B\kern-.05em{\sc i\kern-.025em b}\kern-.08em
    T\kern-.1667em\lower.7ex\hbox{E}\kern-.125emX}}

\usepackage{authblk}
    
\usepackage{xspace}
\newcommand{\eg}{\emph{e.g.}\xspace}    
\newcommand{\ie}{\emph{i.e.}\xspace}

\usepackage{xprintlen}
\usepackage[kerning,spacing]{microtype}
\usepackage{ifthen}
\usepackage[normalem]{ulem} %
\usepackage{xcolor}
\usepackage{amssymb}

\newboolean{showedits}
\setboolean{showedits}{true} %
\ifthenelse{\boolean{showedits}}
{
	\newcommand{\del}[1]{\textcolor{red}{\sout{#1}}} %
}{
	\newcommand{\del}[1]{} %
	
}

\newboolean{showcomments}
\setboolean{showcomments}{true}
\newcommand{\id}[1]{$-$Id: scgPaper.tex 32478 2010-04-29 09:11:32Z oscar $-$}

\ifthenelse{\boolean{showcomments}}
{\newcommand{\nbc}[3]{
 {\colorbox{#3}{\bfseries\sffamily\scriptsize\textcolor{white}{#1}}}
 {\textcolor{#3}{\sf\small$\blacktriangleright$\textit{#2}$\blacktriangleleft$}}}
 }
{\newcommand{\nbc}[3]{}
 \renewcommand{\del}[1]{} %
 }

\definecolor{ibcolor}{rgb}{0.4,0.6,0.2}
\definecolor{clcolour}{rgb}{0.5,0.7,0.9}
\definecolor{accolour}{rgb}{1,0.5,0}
\definecolor{pmcolour}{rgb}{1,0,0}
\definecolor{amethyst}{rgb}{0.6, 0.4, 0.8}

\usepackage{wasysym}

\usepackage{titlesec}
\titlespacing*{\section}{0pt}{2mm}{1mm}
\titlespacing*{\section}{0pt}{2mm}{1mm}
\titlespacing*{\subsection}{0pt}{2mm}{1mm}

\usepackage{setspace}
\usepackage{tikz}

\newcommand{\circled}[1]{{\small\protect\raisebox{0.5pt}{\textcircled{\raisebox{-.2pt}{\textls[-50]{\relsize{-1.5}\phantom{0}\makebox[0pt][c]{#1}\phantom{0}}}}}}\xspace}

\usepackage{url}
\makeatletter
\g@addto@macro{\UrlBreaks}{\UrlOrds}
\makeatother

\newcommand{\todo}[1]{}
\renewcommand{\todo}[1]{{\color{red} TODO: {#1}}}
\newcommand{\fakepara}[1]{\vspace{0mm}\noindent\textbf{#1}\quad}

\definecolor{figurecolor}{RGB}{22,90,220}
\definecolor{citecolor}{RGB}{198,81,19}
\hypersetup{ colorlinks=true, urlcolor=black, linkcolor=figurecolor, citecolor=citecolor, pdfstartview=FitH,
        pdftitle={Aggregate-Driven Trace Visualizations for Performance Debugging (Vision)},
        pdfauthor={}
        }

\usepackage{etoolbox}
\makeatletter
\patchcmd{\maketitle}
	{\@maketitle}
	{\@maketitle\vspace{-5.0em}}
	{}
	{}
\makeatother

\begin{document}

\title{Aggregate-Driven Trace Visualizations for Performance Debugging}

\author[1,2]{Vaastav Anand}
\author[1]{Matheus Stolet}
\author[2]{Thomas Davidson}
\author[1]{Ivan Beschastnikh}
\author[1]{Tamara Munzner}
\author[2]{Jonathan Mace}
\affil[1]{The University of British Columbia}
\affil[2]{Max Planck Institute for Software-Systems}

\date{}

\maketitle

\begin{abstract}

Performance issues in cloud systems are hard to debug. Distributed tracing is a widely adopted approach that gives engineers 
visibility into cloud systems. Existing trace analysis approaches focus on debugging single request correctness
issues but not debugging single request performance issues.
Diagnosing a performance issue in a given request requires comparing the performance of the
offending request with the aggregate performance of typical requests.
Effective and efficient debugging of such issues faces three challenges: (i) identifying the correct aggregate data for diagnosis;
(ii) visualizing the aggregated data; and (iii) efficiently collecting, 
storing, and processing trace data.

We present TraVista, a tool designed for debugging performance issues in a single trace 
that addresses these challenges. TraVista extends the popular single trace Gantt chart visualization with three types of 
aggregate data - metric, temporal, and structure data,
to contextualize the performance of the offending trace across all traces.

\end{abstract}

\section{Introduction}
\label{sec:intro}

Prevalent cloud system designs, like microservices and serverless,
increase the complexity of systems by spreading the execution of a
request across many machines.  This impedes the operators
from diagnosing performance problems, as it requires deducing root
causes from distributed symptoms~\cite{beyer2016site,mace2015pivot}.
Distributed tracing is as an effective way to gain visibility
across distributed
systems~\cite{mace2015pivot,mace2018universal,fonseca2007xtrace}.
Distributed tracing frameworks, like
OpenTelemetry~\cite{opentelemetry}, Jaeger~\cite{jaeger}, and
Zipkin~\cite{zipkin}, are used by most major internet
companies~\cite{kaldor2017canopy,sigelman2010dapper,netflixtracing}.

Distributed tracing tools arose out of a need to understand the
behavior of \emph{individual requests}: identifying the services
invoked, diagnosing problematic requests, and debugging correctness
issues~\cite{fonseca2007xtrace,sigelman2010dapper,macewe}. 
Thus, the user interfaces and visualizations in distributed
tracing tools center on providing an in-depth view into the
execution of a single request.
Diagnosing \emph{system-wide} cloud performance issues remains a
human-driven task~\cite{beyer2016site} as requests must be aggregated
and compared to explain performance anomalies~\cite{shkurographdiffviz}. This is because a request has a
performance anomaly only with respect to \emph{other} requests.

Aggregate analysis of tracing data has been successfully used 
for accomplishing specific tasks where the relevant aggregate analysis task is known \emph{a priori}. For example, modeling
workloads and resource usage~\cite{barham2003magpie,barham2004using,mann2011modeling,thereska2006stardust},
and outlier detection~\cite{kavulya2012draco,wang2012vscope,bailis2016macrobase}.
However, diagnosis of performance issues requires general purpose
aggregation of trace data as a user must evaluate
different hypotheses for the root cause of the issue,
with each hypothesis potentially requiring a different form
of trace aggregation.

Such general purpose aggregation for effective performance debugging is 
non-trivial for three reasons: (i) the aggregate data required is not known a priori;
(ii) designing a visualization tool that effectively supports exploration
of complex traces is non-trivial; and (iii) designing an efficient backend
for collecting, storing, and processing trace data is hard.
Existing trace analysis systems, like Canopy~\cite{kaldor2017canopy} 
have focused primarily on the third challenge and
now can efficiently process billions of traces per day. However, analysis of
this data for performance debugging requires a user to sift through this
large amount of data. This requires building efficient visualization
tools that can support general purpose aggregate queries. This
problem is conceptually and computationally intractable without
having any a priori knowledge about the kind of queries a user might perform.
To this effect, existing visualization tools take a simple approach
by only showing graphs of high level metrics. However, this approach
is not sufficient for performance debugging as that requires
jointly considering aggregate data across the temporal, structural, and metric
dimension of the traces.

To addresses these challenges, we present our initial work on \emph{TraVista}: a tool
for distributed traces designed for diagnosing performance issues.
TraVista builds on the prevalent approach of visualizing a single request
by augmenting this view with information to aid
the user in contextualizing the performance of one request with respect to other requests.
TraVista constrains the possible aggregations to those that are relevant to
spans, events, and relationships present in the trace, instead of considering
all possible aggregations across all traces.
Thus, TraVista builds on the established Gantt chart visualization of a single trace
as it provides a familiar starting point to distributed tracing users and narrows
the scope of aggregation information to three kinds: metric, temporal, and structural
aggregation.

\section{Background \& Motivation}

\begin{figure}
    \centering%
    \includegraphics[width=\linewidth]{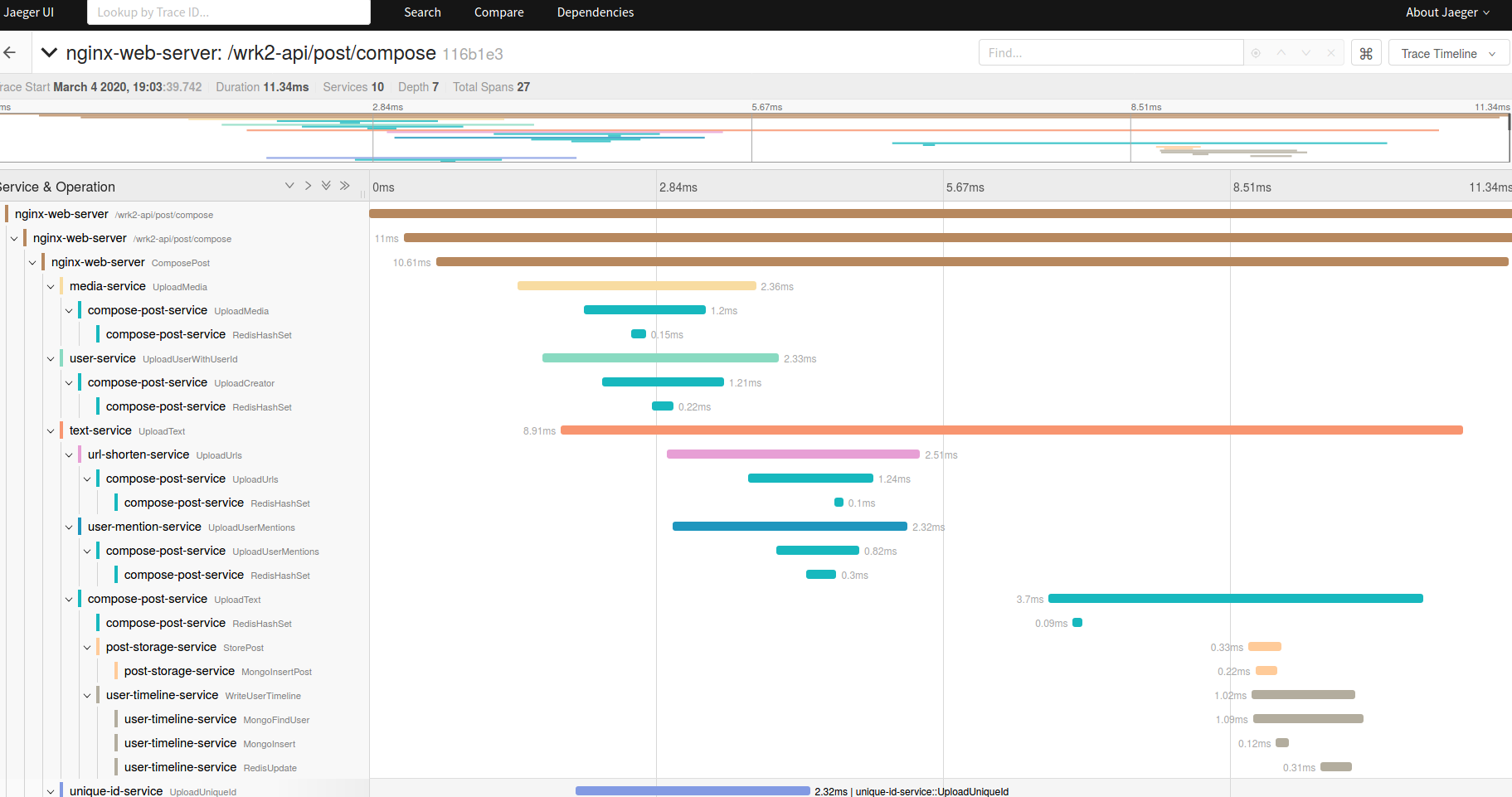}
    \caption{Gantt chart visualization from Jaeger~\cite{jaeger}.}%
    \label{fig:ganttviz}%
    \vspace{-4mm}
\end{figure}

We briefly review the background on distributed tracing and describe three challenges to building trace analysis tools for performance debugging.

\label{sec:background}

\textbf{Trace Structure} \quad A trace is a Directed Acyclic Graph (DAG) of \emph{spans}. A span can be thought
of as a particular task that a system performs to execute a request. The granularity of the task
is user-defined and controlled. A span can represent many things, including a single function execution, 
a single thread execution, or a single operating system process comprised of multiple threads.
Within a span, developers can also add \emph{events}, which are typically unstructured log messages.
Span annotations and events are developer-defined and are connected together with 
\emph{edges}, representing happened-before relationships~\cite{lamport1978time}.
Edges \emph{between} spans typically represent network communication (e.g., the send and receive of RPCs).
Each span records its timing and duration, as well as arbitrary key-value annotations provided by a developer: such as logging a span's arguments.
Each individual trace can be very large, comprising thousands of spans and events~\cite{kaldor2017canopy, las2019sifter, shkurographdiffviz}, and production systems capture traces for millions of requests per day~\cite{kaldor2017canopy}.

Most distributed tracing frameworks represent traces using spans~\cite{jaeger,opentelemetry,sigelman2010dapper}, but some frameworks are based only on events~\cite{fonseca2007xtrace}.  
For event-based frameworks, it is straightforward to group events together into spans (\eg events in the same thread). 
We use the term \emph{task} to refer to both of these concepts. In span-based tracing frameworks a task corresponds to a span and in event-based tracing
frameworks a task corresponds to a collection of events.

\textbf{Gantt chart Visualization} \quad The Gantt chart is the most popular way to represent a trace.
A Gantt chart provides a concise view of all the tasks in a trace, and their respective latencies.
\autoref{fig:ganttviz} shows a Gantt chart screenshot from Jaeger~\cite{jaeger}.
Jaeger provides a minimal trace visualization:
time is represented on the x-axis, each task is represented as a single lane, and the length of the lane encodes the task's duration.
X-Trace~\cite{fonseca2007xtrace} extends this by superimposing events on task lanes using the event's timestamp as its x-axis position.
The visualization also connects events with edges to show the recorded happened-before
relationships~\cite{lamport1978time}. 
Both visualizations help users understand how a request executes in the system,
and for debugging errors in a request.

\subsection{Aggregate Analysis Challenges}
\label{sec:challenges}

We define \textit{aggregate analysis} as analysis of data collected and aggregated from
a set of traces to answer questions about the state, health, and properties of the distributed
system being traced. These questions can be high-level: ``Is the system running correctly?'', or narrowly focused:
 ``Why did a trace have such high latency?''.
We identify three key challenges of building aggregate analysis tools for tracing data.

\textbf{Choosing Aggregate Data} \quad Traces are richly annotated graph-structured data and it is
 not obvious which dimensions or subset of the data is relevant
for finding the root cause of performance anomalies. Pintrace~\cite{pintrace}, the distributed
tracing system at Pinterest, uses aggregate analysis to compare two different groups of traces to narrow
down the root cause of an error. However, the granularity here is coarse as the comparison
is only performed on the distribution of latencies of the traces in each group.
This is only useful for identifying high-level trends.  Early work on trace comparison similarly compared latency 
distributions, but required traces to be structurally isomorphic~\cite{sambasivan2011diagnosing, sambasivan2013visualizing};
this is rare in practice as most traces are structurally unique~\cite{las2019sifter}.

\textbf{Visualizing the data} \quad Building an effective visualization tool is difficult, especially for high-dimensional
data such as trace data. It is not obvious how to present the data to developers for effective analysis or debugging.
The Gantt chart visualization is prevalent for visualizing a single request~\cite{sridharantraceviewwrong,jaeger}.
However, the main drawback of a Gantt chart is its focus on individual requests~\cite{sridharantraceviewwrong}.
It provides no context for whether the trace represents normal or outlier behavior -- a feature intrinsically dependent on \emph{other} traces. This lack of context makes Gantt charts difficult to use for aggregate analyses.
Recent work at Uber~\cite{shkurographdiffviz} compares the structure of incoming traces with a ``good'' set of traces
to find ``bad'' traces, and then visualizes their difference as a directed graph. Canopy~\cite{kaldor2017canopy}
can generate simple graphs of derived metrics from traces. However, none of the existing tools
jointly visualize aggregate structural, metric, and temporal data.
Designing effective visualizations for aggregate analysis continues to elude
the tracing community~\cite{sridharantraceviewwrong}.

\textbf{Data processing system} \quad A key challenge for developing aggregate visualizations is designing an efficient backend
system to support filtering and aggregating trace data across multiple dimensions.
In particular, it is computationally expensive to filter or group traces based on properties of their graph structure (\eg filtering traces that contain a specific sequence of tasks).
Canopy~\cite{kaldor2017canopy} feature extraction is a post-processing step before storing traces.  Canopy supports database queries over these features, but its statistical analysis workflow does not incorporate trace structure.  The authors
acknowledge that inspecting the structure of traces is a separate workflow handled by a different tool and visualization.

\section{TraVista Goals and Design}
\label{sec:travista}

Our goal with TraVista is to help developers working on distributed systems to
diagnose performance issues in traces.
A key intuition behind TraVista is to extend existing single-trace visualizations; in particular, the Gantt chart 
visualization.  Doing this helps address all three challenges listed in~\autoref{sec:challenges}.  First, a Gantt chart is a
well-established visualization that is a familiar starting point for existing distributed tracing users.  Second, this 
narrows the scope of data to be presented, to only the data that is relevant to the selected trace.  This makes it easier for 
us to aggregate relevant features in the dataset to compare the selected trace to other traces in the system, and 
easier to design an efficient backend, since it only needs to support specific kinds of queries.

\subsection{Aggregation Types}
\label{sec:aggregation}

In the context of visualizing a single trace, there are three kinds of aggregated data that we consider
along three different dimensions - tasks, events, and edges. We chose these dimensions because they provide
enough information for us to understand the structural and temporal layout of the traces in the dataset, and the
underlying metrics, whilst allowing an efficient visualization and backend to be built.

\textbf{Metrics} \quad Different systems have different \textbf{performance metrics} to quantify the health of their services.
Common examples include request latency, memory utilization, and throughput, though many systems also define application-level metrics.
Aggregating a metric can mean calculating a statistical quantity or storing the entire distribution. 
Performance metrics can also be broken down using various filters such as a service name, host name, geographical location, etc.
TraVista does metric aggregation for tasks and presents the metric distribution across all traces for each task in
the trace.

\textbf{Temporal Aggregation} \quad At any given point
in time there may be multiple requests being serviced by the system. The information present in a trace is limited to
a single request and does not contain information about concurrently executing requests.
The root cause of outliers can sometimes be concurrent work, \eg high latency can be caused by resource contention.
Zeno~\cite{wu2019zeno} uses a similar idea for performance debugging but solely focuses on identifying queueing delay.
TraVista generalizes this approach by using temporal data to identify other types of resource contention.
Currently, TraVista uses the number of requests being handled by a process executing a task
to approximate the contention during the execution of that task.

\textbf{Structural Aggregation} \quad Traces are rich sources of structural information about how different services
communicate with each other. Structural differences between traces can be very useful in helping users
identify the root cause of an anomaly, \eg occurrence of unlikely sequences of events or edges in a trace
can highlight and explain abnormalities in a trace.
TraVista uses structural aggregation for both edges and events.
To show anomalous events in tasks, it currently shows the frequency of each event occurring in tasks of that type, and
emphasizes events that occur with low frequency. TraVista uses structural information to display the
frequency that a task invokes its child tasks, and emphasizes edges that occur with low frequency.

\begin{figure*}[thb]
\centering%
\includegraphics[width=\textwidth]{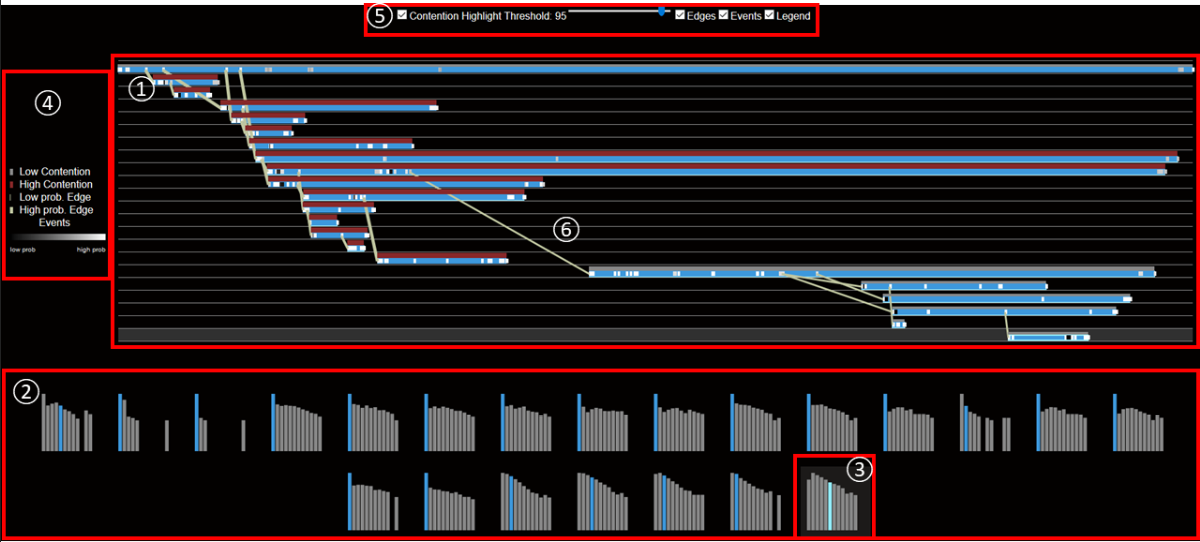}%
\vspace{-2mm}%
\caption{TraVista's Gantt chart visualization extended with aggregate data.}%
\label{fig:travista}%
\vspace{-1mm}%
\end{figure*}

\section{TraVista Interface}
\label{sec:impl}

TraVista comprises of a frontend NodeJS application for visualizing traces and a backend server for storing and querying trace data. To guide users towards potential 
root causes, TraVista makes use of three visualization principles: \emph{eyes beat memory}, the \emph{popout phenomenon}, and 
\emph{detail on demand}~\cite{munzner2014vad}.  We describe these principles inline with our TraVista description below.
For clarity, we describe TraVista's visualization within the context of diagnosing a high latency trace from
an open source X-trace dataset~\cite{anand2019deathstarbenchtraces} consisting of traces from the DeathStar
social network microservice benchmark~\cite{gan2019deathstar}.

\fakepara{Example} \autoref{fig:travista} shows an annotated screenshot of TraVista's trace visualization.  The trace displayed in \autoref{fig:travista} is a high-latency ``ComposePost'' request.  The root cause of this request's high latency is a delay in the Redis cache that stores user timelines.  We will refer to the circled numbers in the figure (\eg \circled{1}) throughout this section.

\fakepara{Gantt Chart}
The starting point of TraVista is the Gantt chart visualization \circled{1}, similar to the visualization shown in \autoref{fig:ganttviz}.
The x-axis represents time, each lane represents a task, and the length of the lane represents the duration of the task.
The Gantt chart gives users an immediate overview of the structure of a
trace and the duration of the different tasks within the trace relative to each other.  TraVista uses the Gantt chart as its starting point because it is a familiar representation for trace data.

\fakepara{Metric Histograms}
We augment the Gantt chart with a list of histograms \circled{2} showing the metric distributions of the different tasks within the trace.  The metric shown is latency as we are trying to diagnose the source of the high latency of the trace.
Each histogram corresponds to one task type (\eg a specific service) and plots its latency distribution across all traces.  Latencies are binned on the x-axis, and y-axis frequencies are plotted in log-scale, since smaller values are intuitively more interesting and must be visually identifiable. As the y-axis is log-scale, any value away from the peak
of the distribution is significantly away from the median of the distribution and is thus potentially anomalous.
The blue highlighted bar in each histogram indicates where the current trace's task falls in the distribution.  Histograms are ordered left-to-right in the same order as the tasks are displayed top-to-bottom in the Gantt chart.  When a user hovers the mouse over a specific histogram \circled{3}, TraVista highlights the corresponding task in the Gantt chart.  In the example, \circled{3} is the ``WriteUserTimeline'' task, which has substantially higher latency than normal.

TraVista displays latency histograms for all tasks simultaneously, adopting the principle of \emph{eyes beat memory}~\cite[Ch. 6.5]{munzner2014vad}.
Using eyes to switch between views that are simultaneously visible has lower cognitive load on the user than consulting memory. 
Reducing cognitive load is important because human working memory is a limited resource, and when its limits are reached, the
user fails to absorb the presented information.

\begin{figure}
\centering%
\includegraphics[width=\columnwidth]{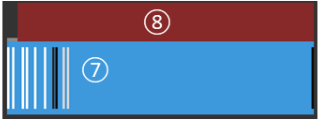}%
\vspace{-3mm}%
\caption{Zoomed-in view of a lane in TraVista's Gantt chart.}%
\label{fig:events}%
\vspace{-1mm}%
\end{figure}

\fakepara{Outlier Events} 
Within the Gantt chart \circled{1}, TraVista augments each lane with two additional sources of information.  The first is information about events that occurred during the task.  \autoref{fig:events} shows
an enlarged view of a lane with this additional information.
TraVista represents events as vertical lines overlaid \circled{7} on the blue rectangles used to represent each task. The x-position
of an event encodes the event's timestamp.

TraVista utilizes the \emph{popout phenomenon}~\cite[Ch. 5.5.4]{munzner2014vad} to aid users in identifying uncommon events by encoding rare events (\ie, events that do not frequently occur in a task) with low luminance (black).  Events that occur frequently are encoded with high luminance (white).  This enables users to quickly identify tasks that have many outlier events.  Users can then dive deeper into the specific rare events. In the highlighted task in \autoref{fig:travista}, two events have low luminance. Both of these correspond to redis updates.

\fakepara{Resource Contention}  The second type of information that TraVista augments the lane in the Gantt chart with is \emph{resource contention}.  We design a novel visualization called \emph{molehills} \circled{8} that allows the user to visualize potential resource 
contention occurring as a result of other tasks executing on the same process at the same time. This data is gathered at a millisecond level 
granularity and linearly scaled based on the maximum value across the trace, where it is plotted as a bar chart on top of every span. 
TraVista
highlights periods where contention is higher than a user-defined percentage threshold \circled{5} in red.

\fakepara{Outlier Edges} TraVista augments the Gantt Chart with edges \circled{6} between spans
which represent happened-before relationships~\cite{lamport1978time}. It utilizes the popout phenomenon~\cite[Ch. 5.5.4]{munzner2014vad} , this time utilizing the size channel rather than the color channel, 
to aid users in identifying uncommon edges by encoding uncommon edges with smaller width and
common edges with bigger width. This allows the users to quickly identify thin edges as the relationships
that are structurally uncommon across the traces.

\fakepara{Detail} 
Since traces can be very detailed, visualizing a trace can lead to significant complexity.  Augmenting a trace with aggregate information increases this complexity.  By default, contention, events, edges, and the legend \circled{4} are disabled, and left as selectable checkboxes allowing users to request \emph{detail on demand}~\cite[Ch. 6.7]{munzner2014vad}.

\section{TraVista Backend}

\begin{figure}
    \centering%
    \includegraphics[width=0.8\columnwidth]{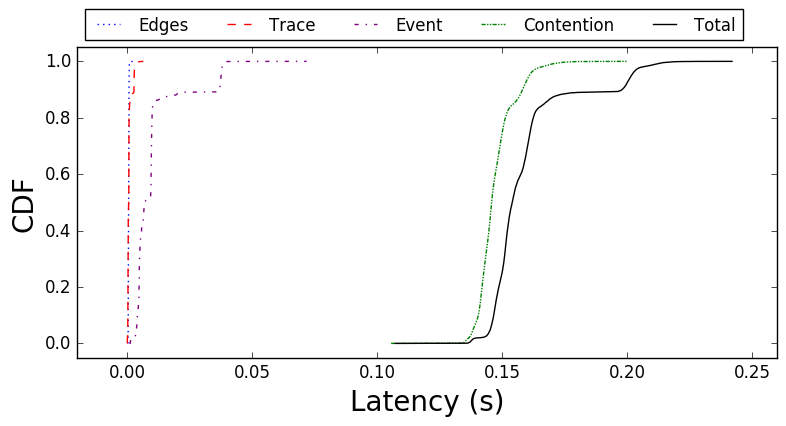}%
    \vspace{-2mm}%
    \caption{Latency breakdown for loading aggregate data.}%
    \label{fig:loadtime}%
    \vspace{-3mm}%
\end{figure}

\begin{figure}
    \vspace{-2mm}%
    \centering%
    \includegraphics[width=0.8\columnwidth]{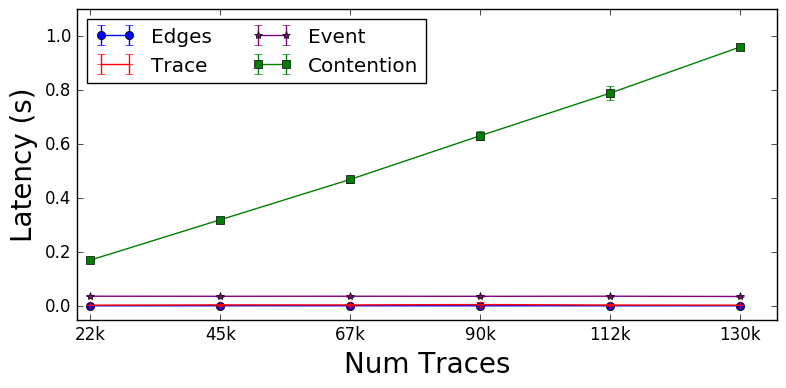}%
    \vspace{-2mm}%
    \caption{Average Latency breakdown for loading a single trace with different number of traces in the database.}%
    \label{fig:scalability}%
    \vspace{-4mm}%
\end{figure}

Our approach for building the backend for TraVista is to follow a simple backend design that allows for efficient querying of aggregated data. Like Canopy~\cite{kaldor2017canopy},
when a new trace is received by the TraVista backend, we extract features and metrics necessary for the visualization.  These 
features are stored in database tables with appropriate indexes for querying. This approach is sufficient for modest datasets like the DeathStarBench~\cite{gan2019deathstar} X-Trace dataset~\cite{anand2019deathstarbenchtraces}. The dataset consists of 22,286 traces (totaling 147,812 tasks and 2,215,250 events).
Pre-processing a trace to extract the relevant features takes around 1ms.
\autoref{fig:loadtime} plots the CDF of the backend latency breakdown for 
loading different kinds of data - raw trace, event aggregate data, edge aggregate data, and contention aggregate data for each trace in the dataset. Backend latency for 
a trace is primarily dependent on the latency of loading contention aggregate data. However, TraVista's current backend approach will not scale over time. \autoref{fig:scalability} shows how the latency for
loading the aggregate data increases linearly with the total number of traces in the database. To perform this experiment,
we generated synthetic copies of the X-Trace dataset and measured the latency breakdown for the biggest trace in the dataset
1000 times. 

Despite the scalability concern with our simple backend, we believe that
bounding the types of aggregations to those relevant to the data in a single trace
can greatly simplify the design of an efficient backend. We believe that
a backend designed to support the performance debugging visualizations
will yield the best results in efficiency and efficacy for performance debugging in the future.
\section{Challenges \& Future Work}
\label{sec:future}

\fakepara{User Study Evaluation}
With TraVista, broad user adoption is one of our core goals, and we believe that achieving this requires
treating usability and visualization as first-class concerns. 
We will collect data about user experiences of using TraVista with
real production systems by running a user-study with production users of Jaeger.

\fakepara{Trace Size} The major drawback of the Gantt Chart visualization is that
it does not scale well with the number of tasks in a trace as it is unable
to fix all the tasks on the screen. ``Scrolling'' is a commonly used tactic
to counter this problem but it is a bad solution as it prevents the user
from getting the full view of a trace.  Moreover, scrolling does not fit
with TraVista's visualization idioms. To fix this, we will
use \emph{detail on demand} to first show a compact picture
of the trace on the screen followed by
using \emph{semantic zooming}~\cite[Ch. 11.5.2]{munzner2014vad}
to allow the user to zoom into different sections of the trace
enabled with TraVista's visualizations. To help the user decide
as to which part to zoom into, we will show the latency histograms
for the top $n$ tasks whose latency deviates from the median.

\fakepara{Aggregate Analysis of Traces}
TraVista is use-case driven, and we followed a `path of least resistance'
by augmenting the Gantt chart visualization.  We do not believe TraVista will solve all aggregate analysis use cases,
and we are interested in exploring other aggregate analysis use cases where this approach will not work.  For example, 
TraVista would not be directly useful for tasks like clustering or trace comparison.

\fakepara{Backends for Complex Data}
Since traces are large and complex graphs, it is infeasible to support real-time queries of arbitrary structural properties.  However, it 
remains unclear to us whether such queries are actually \emph{necessary} for problem diagnosis.  In TraVista, our goal is to develop a 
specialized backend, that only supports the kinds of queries needed by the frontend. Since we know ahead of time the kinds of aggregations we need to do for tasks and events, the 
backend can maintain and update those aggregations with each incoming trace. Eventually, when we need to support more 
complex filtering and grouping, this approach may also not scale.  The solution may require a trade-off between 
 pre-calculated aggregates and on-demand queries.
Our intuition to address this is that our tools do not require perfect results, and can tolerate
a high margin of error. 
Our goal is to support human driven diagnosis, and the ultimate measure of success
is not accurate statistics, but facilitating quicker problem diagnosis.  Techniques such as sub-sampling will be helpful for reducing backend costs.

We believe that TraVista represents a step towards better tools and visualizations for aggregate trace analysis.
TraVista is ongoing work, and in future we will incorporate arbitrary metrics, filtering, and our zooming solution
for traces with large number of microservices.

\bibliographystyle{abbrv}
\bibliography{paper}

\begin{thebibliography}{10}

\bibitem{anand2019deathstarbenchtraces}
V.~Anand and J.~Mace.
\newblock {X-Trace DeathStarBench Dataset}.
\newblock Retrieved October 2019 from
  \url{https://gitlab.mpi-sws.org/cld/trace-datasets/deathstarbench\_traces}.

\bibitem{bailis2016macrobase}
P.~Bailis, E.~Gan, S.~Madden, D.~Narayanan, K.~Rong, and S.~Suri.
\newblock Macrobase: Analytic monitoring for the internet of things.
\newblock {\em arXiv preprint arXiv:1603.00567}, 7, 2016.

\bibitem{barham2004using}
P.~Barham, A.~Donnelly, R.~Isaacs, and R.~Mortier.
\newblock {Using Magpie for Request Extraction and Workload Modelling}.
\newblock In {\em 6th USENIX Symposium on Operating Systems Design and
  Implementation (OSDI '04)}.

\bibitem{barham2003magpie}
P.~Barham, R.~Isaacs, R.~Mortier, and D.~Narayanan.
\newblock Magpie: Online modelling and performance-aware systems.
\newblock In {\em HotOS}, pages 85--90, 2003.

\bibitem{beyer2016site}
B.~Beyer, C.~Jones, J.~Petoff, and N.~R. Murphy.
\newblock {\em Site Reliability Engineering: How Google Runs Production
  Systems}.
\newblock " O'Reilly Media, Inc.", 2016.

\bibitem{netflixtracing}
N.~T. Blog.
\newblock {Lessons from Building Observability Tools at Netflix}.
\newblock Retrieved March 2020 from
  \url{https://netflixtechblog.com/lessons-from-building-observability-tools-at-netflix-7cfafed6ab17}.

\bibitem{opentelemetry}
S.~Flanders.
\newblock {OpenCensus and OpenTracing are now OpenTelemetry}.
\newblock Retrieved March 2020 from
  \url{https://omnition.io/blog/opencensus-and-opentracing-are-now-opentelemetry/}.

\bibitem{shkurographdiffviz}
S.~Flanders and Y.~Shkuro.
\newblock {A Picture is Worth a 1,000 Traces}.
\newblock Retrieved February 2020 from
  \url{https://www.shkuro.com/talks/2019-11-18-a-picture-is-worth-a-thousand-traces/}.

\bibitem{macewe}
R.~Fonseca and J.~Mace.
\newblock {We are Losing Track: a Case for Causal Metadata in Distributed
  Systems}.

\bibitem{fonseca2007xtrace}
R.~Fonseca, G.~Porter, R.~H. Katz, S.~Shenker, and I.~Stoica.
\newblock {X-Trace: A Pervasive Network Tracing Framework}.
\newblock In {\em 4th USENIX Symposium on Networked Systems Design and
  Implementation (NSDI '07)}.

\bibitem{gan2019deathstar}
Y.~Gan, Y.~Zhang, D.~Cheng, A.~Shetty, P.~Rathi, N.~Katarki, A.~Bruno, J.~Hu,
  B.~Ritchken, B.~Jackson, et~al.
\newblock {An Open-Source Benchmark Suite for Microservices and Their
  Hardware-Software Implications for Cloud and Edge Systems}.
\newblock In {\em 24th ACM International Conference on Architectural Support
  for Programming Languages and Operating Systems (ASPLOS '19)}.

\bibitem{jaeger}
{Jaeger: Open Source, End-to-End Distributed Tracing}.
\newblock Retrieved June 2019 from \url{https://www.jaegertracing.io/}.

\bibitem{kaldor2017canopy}
J.~Kaldor, J.~Mace, M.~Bejda, E.~Gao, W.~Kuropatwa, J.~O'Neill, K.~W. Ong,
  B.~Schaller, P.~Shan, B.~Viscomi, V.~Vekataraman, K.~Veeraraghavan, and Y.~J.
  Song.
\newblock {Canopy: An End-to-End Performance Tracing And Analysis System}.
\newblock In {\em 26th ACM Symposium on Operating Systems Principles (SOSP
  '17)}.

\bibitem{pintrace}
S.~Karumuri.
\newblock {Distributed Tracing at Pinterest with New Open-Source Tools}.
\newblock Retrieved March 2020 from
  \url{https://medium.com/pinterest-engineering/distributed-tracing-at-pinterest-with-new-open-source-tools-a4f8a5562f6b}.

\bibitem{kavulya2012draco}
S.~P. Kavulya, S.~Daniels, K.~Joshi, M.~Hiltunen, R.~Gandhi, and P.~Narasimhan.
\newblock Draco: Statistical diagnosis of chronic problems in large distributed
  systems.
\newblock In {\em IEEE/IFIP International Conference on Dependable Systems and
  Networks (DSN 2012)}, pages 1--12. IEEE, 2012.

\bibitem{lamport1978time}
L.~Lamport.
\newblock {Time, Clocks, and the Ordering of Events in a Distributed System}.
\newblock {\em Communications of the ACM}, 21(7):558--565, 1978.

\bibitem{las2019sifter}
P.~Las-Casas, G.~Papakerashvili, V.~Anand, and J.~Mace.
\newblock {Sifter: Scalable Sampling for Distributed Traces, without Feature
  Engineering}.
\newblock In {\em 11th ACM Symposium on Cloud Computing (SOCC '19)}.

\bibitem{mace2018universal}
J.~Mace and R.~Fonseca.
\newblock {Universal Context Propagation for Distributed System
  Instrumentation}.
\newblock In {\em 13th ACM European Conference on Computer Systems (EuroSys
  '18)}.

\bibitem{mace2015pivot}
J.~Mace, R.~Roelke, and R.~Fonseca.
\newblock {Pivot Tracing: Dynamic Causal Monitoring for Distributed Systems}.
\newblock In {\em 25th ACM Symposium on Operating Systems Principles (SOSP
  '15)}.

\bibitem{mann2011modeling}
G.~Mann, M.~Sandler, D.~Krushevskaja, S.~Guha, and E.~Even-Dar.
\newblock {Modeling the Parallel Execution of Black-Box Services}.
\newblock In {\em 3rd USENIX Workshop on Hot Topics in Cloud Computing
  (HotCloud '11)}.

\bibitem{munzner2014vad}
T.~Munzner.
\newblock {\em Visualization Analysis and Design}.
\newblock CRC Press, 2014.

\bibitem{sambasivan2013visualizing}
R.~R. Sambasivan, I.~Shafer, M.~L. Mazurek, and G.~R. Ganger.
\newblock {Visualizing Request-Flow Comparison to Aid Performance Diagnosis in
  Distributed Systems}.
\newblock {\em IEEE Transactions on Visualization and Computer Graphics},
  19(12):2466--2475, 2013.

\bibitem{sambasivan2011diagnosing}
R.~R. Sambasivan, A.~X. Zheng, M.~De~Rosa, E.~Krevat, S.~Whitman, M.~Stroucken,
  W.~Wang, L.~Xu, and G.~R. Ganger.
\newblock {Diagnosing Performance Changes by Comparing Request Flows}.
\newblock In {\em 8th USENIX Symposium on Networked Systems Design and
  Implementation (NSDI '11)}.

\bibitem{sigelman2010dapper}
B.~H. Sigelman, L.~A. Barroso, M.~Burrows, P.~Stephenson, M.~Plakal, D.~Beaver,
  S.~Jaspan, and C.~Shanbhag.
\newblock {Dapper, a Large-Scale Distributed Systems Tracing Infrastructure}.
\newblock {Technical Report}, Google, 2010.

\bibitem{sridharantraceviewwrong}
C.~Sridharan.
\newblock {Distributed Tracing: We've Been Doing It Wrong}.
\newblock Retrieved February 2020 from
  \url{https://medium.com/@copyconstruct/distributed-tracing-weve-been-doing-it-wrong-39fc92a857df}.

\bibitem{thereska2006stardust}
E.~Thereska, B.~Salmon, J.~Strunk, M.~Wachs, M.~Abd-El-Malek, J.~Lopez, and
  G.~R. Ganger.
\newblock {Stardust: Tracking Activity in a Distributed Storage System}.
\newblock In {\em 2006 ACM International Conference on Measurement and Modeling
  of Computer Systems (SIGMETRICS '06)}.

\bibitem{zipkin}
{Zipkin}.
\newblock Retrieved July 2017 from \url{http://zipkin.io/}.

\bibitem{wang2012vscope}
C.~Wang, I.~A. Rayan, G.~Eisenhauer, K.~Schwan, V.~Talwar, M.~Wolf, and
  C.~Huneycutt.
\newblock Vscope: middleware for troubleshooting time-sensitive data center
  applications.
\newblock In {\em ACM/IFIP/USENIX International Conference on Distributed
  Systems Platforms and Open Distributed Processing}, pages 121--141. Springer,
  2012.

\bibitem{wu2019zeno}
Y.~Wu, A.~Chen, and L.~T.~X. Phan.
\newblock {Zeno: Diagnosing Performance Problems with Temporal Provenance}.
\newblock In {\em 16th USENIX Conference on Networked Systems Design and
  Implementation (NSDI '19)}.

\end{thebibliography}

\end{document}